\documentclass[showpacs,nofootinbib,aps,preprintnumbers,preprint,
  floatfix,superscriptaddress]{revtex4}
\usepackage{graphicx} 
\usepackage{epsfig}
\usepackage{psfrag}
\usepackage{dcolumn} 
\usepackage{bm} 

\def\gsim{\lower0.5ex\hbox{$\:\buildrel >\over\sim\:$}}
\def\lsim{\lower0.5ex\hbox{$\:\buildrel <\over\sim\:$}}

\newcommand{\bea}{\begin{eqnarray}}
\newcommand{\eea}{\end{eqnarray}}

\begin{document}

\preprint{HIP-2005-05/TH}
\preprint{hep-ph/0502100}

\title{Phenomenology of non-universal gaugino masses in supersymmetric
  grand unified theories}

\author{Katri Huitu}
\email{Katri.Huitu@helsinki.fi}
\affiliation{High Energy Physics Division, Department of Physical
Sciences, P.O. Box 64, FIN-00014 University of Helsinki, Finland}
\affiliation{Helsinki Institute of Physics,
P.O. Box 64, FIN-00014 University of Helsinki, Finland}
\author{Jari Laamanen}
\email{jalaaman@pcu.helsinki.fi}
\affiliation{High Energy Physics Division, Department of Physical
Sciences, P.O. Box 64, FIN-00014 University of Helsinki, Finland}
\affiliation{Helsinki Institute of Physics,
P.O. Box 64, FIN-00014 University of Helsinki, Finland}
\author{Pran N. Pandita}
\email{ppandita@nehu.ac.in}
\affiliation{Department of Physics, North-Eastern Hill University, 
  Shillong 793 022, India} 
\author{Sourov Roy}
\email{roy@pcu.helsinki.fi}
\affiliation{Helsinki Institute of Physics,
P.O. Box 64, FIN-00014 University of Helsinki, Finland}

\received{\today}

\pacs{12.60.Jv, 11.30.Er, 14.80.Ly} 

\begin{abstract} \vspace*{10pt}
Grand unified theories can lead to non-universal boundary conditions
for the gaugino masses at the unification scale.  We consider the
implications of such non-universal boundary conditions for the
composition of the lightest neutralino as well as for the upper bound
on its mass in the simplest supersymmetric grand unified theory based
on the $SU(5)$ gauge group.  We derive sum rules for neutralino and
chargino masses in different representations of $SU(5)$ which lead
to different non-universal boundary conditions for the gaugino masses
at the unification scale. We also consider the phenomenological
implications of the non-universal gaugino masses arising from a grand
unified theory in the context of Large Hadron Collider.  In particular
we investigate the detection of heavy neutral Higgs bosons $H^0,\;A^0$
from $H^0,\;A^0 \to \tilde{\chi}_2^0\tilde{\chi}_2^0\to 4l $, and
study the possibilities of detecting the neutral Higgs bosons in
cascade decays, including the decays $\tilde{\chi}_2^0\to h^0 (H^0,
A^0)\tilde{\chi}_1^0\to b\bar b \tilde{\chi}_1^0$.
\end{abstract}

\maketitle

\section{Introduction}
Supersymmetry is at present an attractive framework in which the Higgs
sector of the Standard Model~(SM), so crucial for its consistency, is
technically natural.  It is widely expected that some of the
supersymmetric partners of the SM particles will be produced at the
CERN Large Hadron Collider~(LHC) which is going to start operation in
a few years time.  In the experimental search for supersymmetry~(SUSY)
the lightest supersymmetric particle will play a crucial role since
the heavier supersymmetric particles will decay into it.  In SUSY
models with R-parity conservation, the lightest supersymmetric
particle is absolutely stable. The lightest supersymmetric particle is
constrained to be a weakly interacting neutral particle~\cite{WIMP}.

In most of the supersymmetric models the lightest
neutralino~$(\tilde\chi_1^0)$, which is typically an admixture of
gauginos and higgsinos, is the lightest supersymmetric
particle~(LSP). Such an LSP is a good candidate for a particle dark
matter~\cite{DM}. From the point of view of experimental discovery of
supersymmetry at a collider like the LHC, the LSP is the final product
of the cascade decay of a SUSY particle.  In this work we will assume
that the LSP is the lightest neutralino, and that it escapes the
collider experiments undetected. The cascade chain will typically also
contain other neutralinos~$(\tilde\chi^0_j,~ j = 2, 3, 4)$ as well as
charginos~~$(\tilde\chi^{\pm}_i,~ i = 1, 2)$.  The charginos are an
admixture of charged gauginos and charged higgsinos.  The composition
and mass of the neutralinos and charginos will play a key role in the
search for supersymmetric particles.  These properties determine also
the time-scale of their decays.  The mass patterns of the neutralinos
in models with different particle content, or with specific SUSY
breaking patterns were considered in some detail in~\cite{HLP,Pran}.

Although most of the phenomenological studies involving neutralinos
and charginos have been performed with universal gaugino masses at the
grand unification scale, there is no compelling theoretical reason for
such a choice.  Gaugino masses follow from higher dimensional
interaction terms which involve gauginos and auxiliary parts of chiral
superfields in a given supersymmetric model.  Assuming an $SU(5)$
grand unified theory (GUT) model, the auxiliary part of a chiral
superfield in these higher dimensional terms can be in the
representation {\bf 1}, {\bf 24}, {\bf 75}, or {\bf 200}, or some
combination of these, of the underlying $SU(5)$ gauge group. If the
auxiliary field of one of the $SU(5)$ nonsinglet chiral superfields
obtains a vacuum expectation value (VEV), then the gaugino masses are
not universal at the grand unification scale. Moreover, nonuniversal
soft supersymmetry breaking masses, like gaugino masses, are a
necessary feature in some of the supersymmetric models, e.g. in
anomaly mediated supersymmetry breaking models the gaugino masses are
not unified \cite{amsb}.

As indicated above, the phenomenology of supersymmetric models depends
crucially on the composition of neutralinos and charginos.  Thus, it
is important to investigate the changes in the experimental signals
for supersymmetry with the changes in the composition of neutralinos
and charginos that may arise because of the changes in the underlying
boundary conditions at the grand unification scale, or when the
underlying supersymmetric model is changed. The implications of
nonuniversal gaugino masses has been considered in a number of works,
e.g. in a study of constraints arising from experimental
measurements~\cite{coll,constr,DMM}, and in the context of
supersymmetric dark matter~\cite{DM2,pnath}.  In~\cite{DMM}, the
decays of the second lightest neutralino were studied in the context
of nonuniversal gaugino masses.
 
In this paper we shall study the implications of the nonuniversal
gaugino masses for the phenomenology of neutral Higgs bosons.  It has
been known for quite some time that the cascade decays of the SUSY
particles may be a major source of the Higgs bosons
\cite{HPSSY,DDGM,BBTW}: the copiously produced strongly interacting
particles can cascade decay to the Higgs bosons.  In addition to the
obvious interest in producing the Higgs bosons, it has been realized
that this method of producing the Higgs bosons does not depend on the
value of $\tan\beta$.  Thus, this method of producing Higgs bosons may
help to cover a larger parameter space as compared to the more
conventional methods of studying the Higgs sector of supersymmetric
models, including also the heavier Higgs bosons. The gauginos also
play an important role in the decays of Higgs bosons when they are
kinematically allowed to decay to the second lightest neutralino pair,
which in turn may decay to the lightest neutralinos and two leptons
\cite{BBKT}. Such a signal seems to be relatively easy to discover at
the LHC \cite{cms1,cms2}. We note here that Higgs boson production via 
cascade decays and detection via Higgs decay to neutralinos has been 
studied in CMS detector simulations at LHC \cite{cms1,cms2,cms3} in 
the case of minimal supersymmetric
standard model~(MSSM) with universal gaugino masses.  
Here we study the Higgs production and decay when gaugino masses
are nonuniversal.

The plan of the paper is as follows. In Section II we consider in
detail the nonuniversality of gaugino masses as it arises in $SU(5)$
supersymmetric grand unified theory. In this Section we consider
analytically the implications of such a nonuniversality for neutralino
and chargino masses. We derive sum rules involving the neutralino and
chargino squared masses when the supersymmetry breaking gaugino masses
are nonuniversal. In Section III we consider the phenomenology of
Higgs bosons when the gaugino masses are nonuniversal. In this Section
we consider Higgs decays to heavier neutralinos which then
cascade into the lightest neutralino and leptons.  In Section IV we
calculate the production of squark and gluino pairs in a particular
scenario where the gluinos are heavier than squarks, and then study
the cascade decays of the squarks into Higgs bosons. We conclude our
paper with a summary in Section V.


\section{ Nonuniversal gaugino masses in supersymmetric $SU(5)$}
The masses and the compositions of neutralinos and charginos are
determined by the soft supersymmetry breaking gaugino masses $M_1$,
$M_2$, and $M_3$, corresponding to $U(1)$, $SU(2)$, and $SU(3)$ gauge
groups, respectively, the supersymmetric Higgs mixing parameter $\mu$,
and the ratio of the vacuum expectation values of the two
neutral Higgs bosons $H_1^0$ and $H_2^0$, $\langle H_2^0\rangle
/\langle H_1^0\rangle =\tan\beta$. In the simplest supersymmetric
models with universal gaugino masses, $M_1$, $M_2$, and $M_3$ are
taken to be equal at the grand unified scale. However, in
supersymmetric theories with an underlying grand unified gauge group,
the gaugino masses need not be equal at the GUT scale. In this Section
we consider the nonuniversality of gaugino masses as it arises in the
simplest of the supersymmetric grand unified theories, namely
supersymmetric $SU(5)$ grand unified theory, and its implications.

In grand unified supersymmetric models, including $SU(5)$
grand unified models, non-universal gaugino masses are generated by a 
non-singlet chiral superfield $\Phi^n$ that appears linearly in the gauge 
kinetic function $f(\Phi)$~(the chiral superfields $\Phi$  are classified 
into a set of gauge singlet superfields $\Phi^s$, and gauge nonsinglet
superfields $\Phi^n$, respectively under the grand unified group), 
which is an analytic function of the 
chiral superfields $\Phi$ in the theory \cite{CFGP}.
If the auxiliary part $F_\Phi$ of a chiral superfield $\Phi$  in 
$f(\Phi)$ gets a VEV, then gaugino masses arise from the coupling of 
$f(\Phi)$ with the field strength superfield $W^a$. The Lagrangian
for the coupling of gauge kinetic function to the gauge field strength
is written as
\bea 
{\cal L}_{g.k.} \; = \;
\int d^2\theta f_{ab}(\Phi) W^{a}W^{b}
+h.c.,
\label{gk}
\eea
where $a$ and $b$ are gauge group indices, and repeated indices are summed 
over. The gauge kinetic function $f_{ab}(\Phi)$ is 
\bea
f_{ab}(\Phi) & = & f_0(\Phi^s)\delta_{ab} 
+ \sum_n f_n(\Phi^s){\Phi_{ab}^n\over M_P} + \cdot \cdot \cdot, 
\eea
where as indicated above the $\Phi^s$ and the  $\Phi^n$ are the singlet and 
the non-singlet chiral superfields, respectively. Here
$f_0(\Phi^s)$ and $f_n(\Phi^s)$ are functions of gauge singlet
superfields $\Phi^s$, and $M_P$ is some large scale.  When $F_\Phi$ 
gets a VEV $\langle F_\Phi \rangle$, the interaction~(\ref{gk})
gives rise to gaugino masses:  
\bea
{\cal L}_{g.k.} \; \supset \;
{{{\langle F_\Phi \rangle}_{ab}} \over {M_P}}
\lambda^a \lambda^b +h.c., 
\eea
where $\lambda^{a,b}$ are gaugino fields. Note that we denote by
$\lambda_1$, $\lambda_2$ and $\lambda_3$ the $U(1)$, $SU(2)$ and $SU(3)$ 
gauginos, respectively. Since the gauginos belong to the adjoint 
representation of $SU(5)$, $\Phi$ and $F_\Phi$ can belong to any of the 
following representations appearing in the symmetric product of the 
two {\bf 24} dimensional representations of $SU(5)$: 
\bea 
({\bf 24 \otimes 24})_{Symm} = {\bf 1 \oplus 24 \oplus 75 \oplus 200}.
\label{product}
\eea 
In the minimal, and the simplest, case $\Phi$ and $F_\Phi$ are
assumed to be in the singlet representation of $SU(5)$, which implies
equal gaugino masses at the GUT scale.  However, as is clear from the
decomposition (\ref{product}), $\Phi$
can belong to any of the non-singlet representations 
{\bf 24}, {\bf 75},  and {\bf 200} of $SU(5)$, in which case
these gaugino masses are unequal but related to one another via the
representation invariants~\cite{Enqvist_et_al}.  In Table~\ref{tab1}
we show the ratios of resulting gaugino masses at tree-level as they
arise when $F_{\Phi}$ belongs to various representations of $SU(5)$.
For definiteness, we shall study the case of each representation
separately, although an arbitrary combination of these is obviously
also allowed.

\medskip

\begin{table}[htb]
  \caption{\label{tab1} Ratios of the gaugino masses at the GUT scale
    in the normalization ${M_3}(GUT)$ = 1, and at the electroweak
    scale in the normalization ${M_3}(EW)$ = 1 at the 1-loop level.}
  \centering
  \begin{tabular}{c||ccc||ccc}
   \hline \hline
    $F_\Phi$ & $M_1^G$ & $M_2^G$ & $M_3^G$ & 
    $M_1^{EW}$ & $M_2^{EW}$ & $M_3^{EW}$
    \\ \hline \hline {\bf 1} & 1 & 1
    & 1 & 0.14 & 0.29 & 1
    \\ {\bf 24} & -0.5 & -1.5 & 1 & -0.07 & -0.43 & 1
    \\ {\bf 75} & -5 & 3 & 1 & -0.72 & 0.87 & 1
    \\ {\bf 200} & 10 & 2 & 1 &1.44 & 0.58 & 1
    \\ \hline \hline
  \end{tabular}
\end{table}

\noindent These results are consistent with the unification of gauge
couplings \bea \alpha^G_3 = \alpha^G_2 = \alpha^G_1 = \alpha^G
(\approx 1/25), \eea at the grand unification scale, where we have
neglected the contribution of nonuniversality to the gauge
couplings. Such contributions have little effect on the
phenomenological aspects that we are interested in this paper.
Because of the renormalization group (RG) evolution we have at any
scale (at the one-loop level)~\cite{martin-ramond}

\bea {{M_i (t)} \over {\alpha_i(t)}} = {{{M_i}({\rm GUT})} \over
{{\alpha_i}({\rm GUT})}}.  \eea

\noindent Thus,  at any scale we have
\bea
{M_1} = {\frac 5 3}{{\alpha} \over {\cos^2{\theta_W}}}
\left({{{M_1}({\rm GUT})} \over {{\alpha_1}({\rm GUT})}}\right),\;\;
{M_2} = {{\alpha} \over {\sin^2{\theta_W}}} 
\left({{{M_2}({\rm GUT})} \over {{\alpha_2}({\rm GUT})}}\right),\;\;
{M_3} = {\alpha_3} \left({{{M_3}({\rm GUT})} \over 
{{\alpha_3}({\rm GUT})}}\right).\nonumber\\
\eea

\noindent For the {\bf 24} dimensional representation of $SU(5)$, we
then have

\bea
{{M_1} \over {M_3}} = -\frac 12 \left({\frac 5 3}{{\alpha} \over
{\cos^2{\theta_W}}}\right) \left({1 \over {\alpha_3}}\right),\;\;
{{M_2} \over {M_3}} = -\frac 32\left({{\alpha} \over
{\sin^2{\theta_W}}}\right) \left({1 \over {\alpha_3}}\right).
\eea

\noindent Similarly, for the {\bf 75} dimensional representation of
$SU(5)$, we have

\bea
{{M_1} \over {M_3}} = -5 \left({\frac 5 3}{{\alpha} \over
{\cos^2{\theta_W}}}\right) \left({1 \over {\alpha_3}}\right),\;\;
{{M_2} \over {M_3}} = 3\left({{\alpha} \over
{\sin^2{\theta_W}}}\right) \left({1 \over {\alpha_3}}\right),
\eea

\noindent and for the {\bf 200} dimensional representation of $SU(5)$ we have
\bea
{{M_1} \over {M_3}} = 10 \left({\frac 5 3}{{\alpha} \over
{\cos^2{\theta_W}}}\right) \left({1 \over {\alpha_3}}\right),\;\;
{{M_2} \over {M_3}} = 2 \left({{\alpha} \over
{\sin^2{\theta_W}}}\right) \left({1 \over {\alpha_3}}\right).  \eea We
can scale down these results to the the electroweak scale by using the
relevant renormalization group equations. At the electroweak scale we
have the result $M_1<M_2$ for the singlet representation,
$|M_1|<|M_2|$ for the {\bf 24} and {\bf 75} representation, and
$M_1>M_2$ for {\bf 200} dimensional representation of $SU(5)$,
respectively. The approximate values for the soft gaugino masses at
the weak scale, $M_i(EW)$ are shown in Table~\ref{tab1}. These are calculated
using one loop RG-equations for the gaugino masses and the gauge
couplings. Two-loop effect is to increase the $M_1/M_2$-ratio
slightly.
\begin{figure}[t]
\leavevmode
\begin{center}
\mbox{\epsfxsize=7.truecm\epsfysize=5.truecm\epsffile{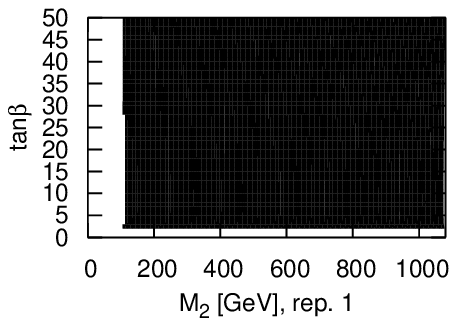}}
\mbox{\epsfxsize=7.truecm\epsfysize=5.truecm\epsffile{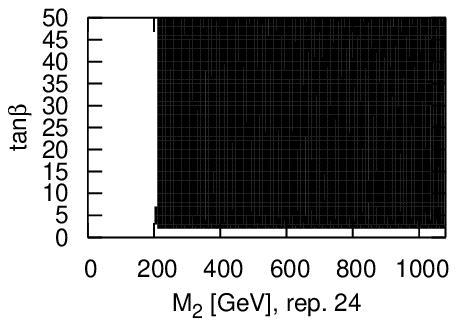}}\\
\mbox{\epsfxsize=7.truecm\epsfysize=5.truecm\epsffile{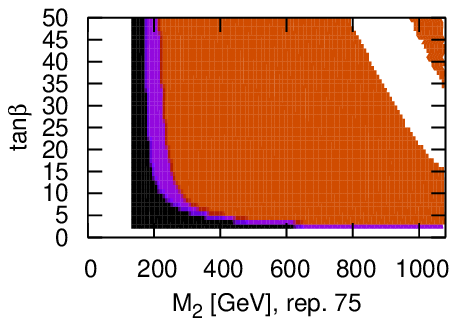}}
\mbox{\epsfxsize=7.truecm\epsfysize=5.truecm\epsffile{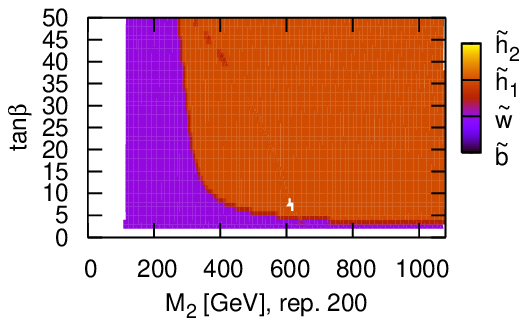}}
\end{center}
\caption{\label{composition} Main component of the lightest neutralino
  in different representations of $SU(5)$ that arise in the
  product~(\ref{product}) for a common universal scalar mass
  $m_0=1$ TeV given at the GUT scale.  The value of $M_2$ is
  calculated and plotted at the electroweak scale.  } 
\end{figure}

In Fig.~\ref{composition} we have shown the dominant component of the
lightest neutralino~(LSP) for the four representations as a function
of $\tan\beta$ and $M_2(EW)$ for the value of soft supersymmetry breaking
scalar mass $m_0(GUT)=1$ TeV.  The values of $\mu$ used in the computations
were determined by requiring the radiative electroweak symmetry
breaking at the relevant scale. The universal trilinear coupling $A_0$
was set to zero at the GUT scale and the sign of the Higgs mixing
parameter $\mu$ was set to +1, but the choice of the sign is not
crucial to the composition of the lightest neutralino. The scan was
done using the program SOFTSUSY \cite{SOFTSUSY} that uses two loop 
RG $\beta$-functions for the relevant parameters.

For the case of the singlet representation, the dominant component is
always the bino, as expected.  This is also true for the {\bf 24}
dimensional representation of $SU(5)$.  For the singlet case the
experimental mass limit of the lighter chargino ($m_{\tilde{\chi}^\pm_1} >
103$ GeV if $m_{\tilde{\nu}}>200$ GeV, $m_{\tilde{\chi}^\pm_1} > 45$
GeV if $m_{\tilde{\nu}}<200$ GeV \cite{PRD}) restricts the lower end
of the $M_2$ range. In the {\bf 24} dimensional representation the
lower end of the $M_2$ range is restricted by the lightest neutralino
mass limit $m_{\tilde{\chi}^0_1} > 36$ GeV \cite{PRD}.

For the {\bf 75} dimensional representation of $SU(5)$, we have
several possibilities.  For the value of the soft parameter $m_0=1$
TeV, one has a bino LSP for small values of $M_2$, a wino LSP for
slightly larger values of $M_2$, and a higgsino LSP for $M_2\gsim 300$
GeV, all for a value of $\tan\beta\gsim 10$.  In the case of {\bf 75}
dimensional representation there exists a band of discontinuity in the
$(M_2,\tan\beta)$-parameter space. For these values of parameters the
lighter chargino mass becomes too light.  The lower end of the $M_2$
range is restricted in this case by the experimental limit on the
gluino mass.

As seen in Fig.~\ref{composition}, for the {\bf 200} dimensional
representation the LSP is either a wino or a higgsino, depending on
the values of $M_2$ and $\tan\beta$.  Here, as in the singlet case,
the experimental lightest chargino mass limit restricts the lower end
of the $M_2$ range.  Also in the {\bf 200} dimensional representation
there is a small region around $7\lsim \tan\beta \lsim 8$,
$610~\textrm{GeV}< M_2(\textrm{EW}) < 620~\textrm{GeV}$, where the
experimental mass limits of charginos (and also neutralinos) are not
met.
\begin{figure}[t]
  \psfrag{N1}{$m_{\tilde \chi_{1}^0}$ \scriptsize[GeV]}
  \psfrag{N2}{$m_{\tilde \chi_{2}^0}$ \scriptsize[GeV]}
  \psfrag{mg}{$m_{\tilde g}$ \scriptsize[GeV]} %
  \centering
  \includegraphics[width=7.5cm]{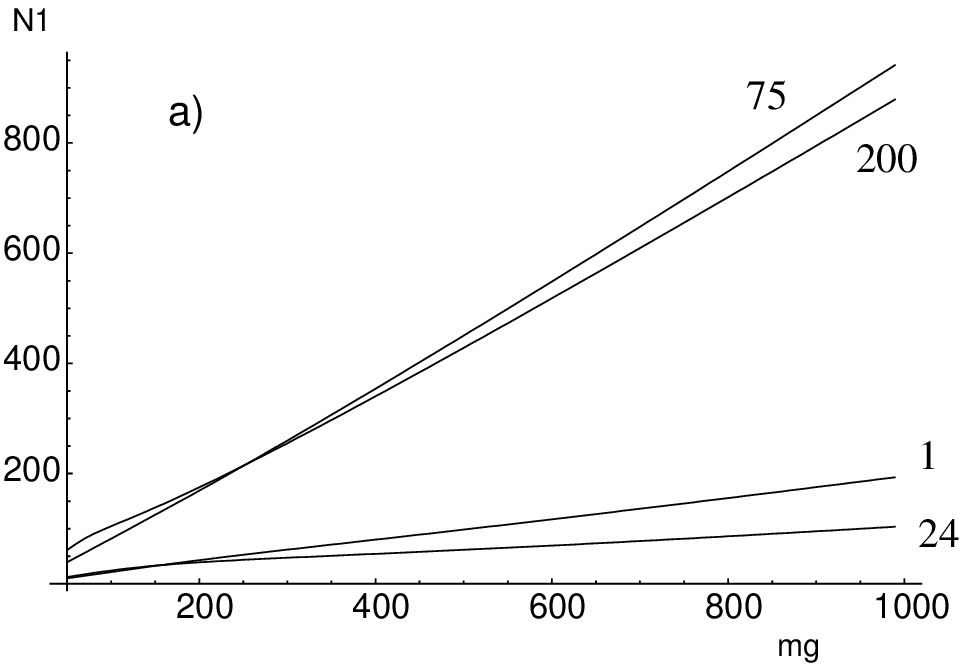} \centering
  \includegraphics[width=7.5cm]{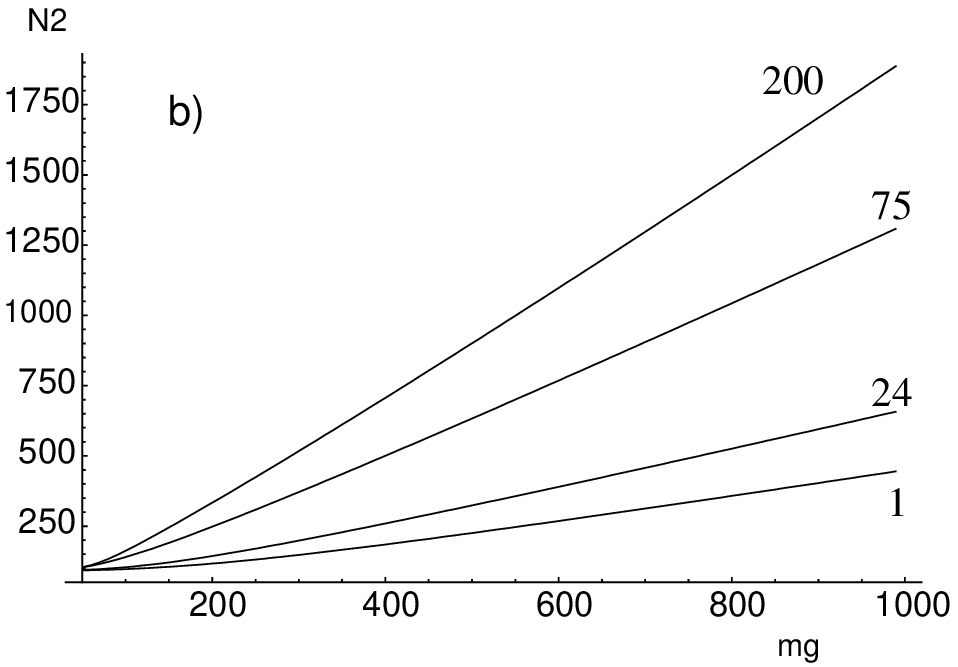}
  \caption{\label{upper1}The upper bound for (a) the lightest
    neutralino mass and (b) for the second lightest neutralino mass
    for different representations that arise in (\ref{product}).}
\end{figure}

We recall here that there is a general upper bound on the mass of the
lightest neutralino~$(\tilde{\chi}^0_1)$ that follows from the
structure of the neutralino mass matrix~\cite{Pran,HLP}. This upper
bound can be written as \bea M_{\tilde{\chi}^0_1}^2 \le \frac 12
\left( M_1^2 + M_2^2 +M_Z^2 - \sqrt{(M_1^2 - M_2^2)^2 +M_Z^4 -2 (M_1^2
- M_2^2)M_Z^2\cos 2\theta_W }\right) .
\label{bound1}
\eea In Fig.~\ref{upper1} we plot this upper bound for the lightest
neutralino mass for the four different representations of $SU(5)$ that
we have considered in this paper.  From Fig.~\ref{upper1} (a) we see
that the large coefficients in the Table~\ref{tab1} result in large
differences in the upper bound on the mass of the lightest neutralino
for the four different representations in (\ref{product}).  Similarly,
as discussed in \cite{HLP}, an upper bound can be obtained for the
second lightest neutralino.  This upper bound for $\tilde{\chi}_2^0$
is shown in Fig.~\ref{upper1} (b). The gaugino masses here are
calculated in the next-to-leading order 
(see e.g.~\cite{HLP}).
\begin{figure}[t]
  \leavevmode \centering
  \includegraphics[height=7.5cm]{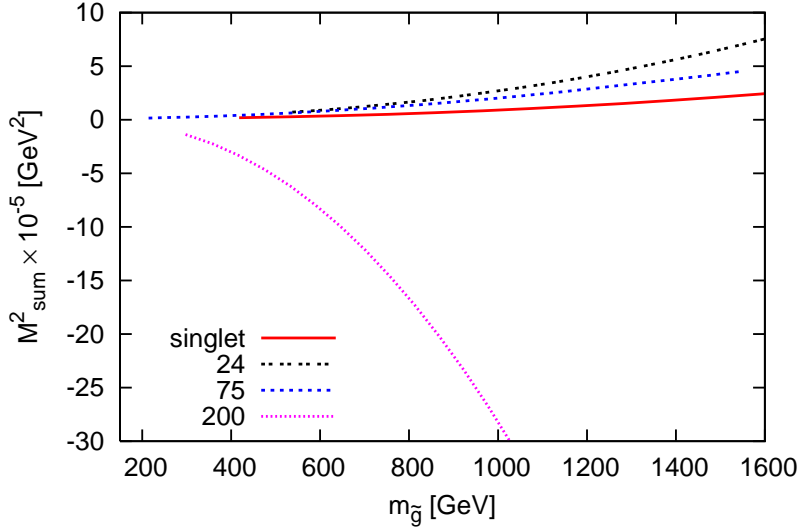}
  \caption{\label{sumrule} The sum rule $2\sum m_{\tilde\chi_i^\pm}^2
  - \sum m_{\tilde\chi_i^0}^2$ as a function of the gluino mass.}
\end{figure}

{}In order to study analytically the implications of the nonuniversal
gaugino masses on the neutralino and chargino mass spectrum, we
consider the trace of the neutralino and chargino mass squared
matrices.  From the trace of these matrices, we can calculate the
average mass squared difference of the charginos and neutralinos. This
mass squared difference depends only on the physical masses, and not
on the Higgs(ino) mass parameter $\mu$ or the ratio of VEV's,
$\tan\beta$~\cite{martin-ramond}.  For the four different
representations of $SU(5)$ which arise in (\ref{product}), 
we find at the tree-level the sum rules
\bea
M^2_{sum} = 2(M^2_{\tilde{\chi}_1^\pm}+M^2_{\tilde{\chi}_2^\pm})&-&
(M^2_{\tilde{\chi}_1^0}+M^2_{\tilde{\chi}_2^0}+M^2_{\tilde{\chi}_3^0}+
M^2_{\tilde{\chi}_4^0})\nonumber\\
&&=(\alpha_2^2-\alpha_1^2)\frac{M_{\tilde g}^2}{\alpha_3^2} +
4 m_W^2 -2m_Z^2, \;\;{\rm for\; {\bf 1}},\\
&&=(\frac 94 \alpha_2^2-\frac 14\alpha_1^2)\frac{M_{\tilde g}^2}{\alpha_3^2} +
4 m_W^2 -2m_Z^2, \;\;{\rm for\; {\bf 24}},\\
&&=(9\alpha_2^2-25\alpha_1^2)\frac{M_{\tilde g}^2}{\alpha_3^2} +
4 m_W^2 -2m_Z^2, \;\;{\rm for\; {\bf 75}},\\
&&=(4\alpha_2^2-100\alpha_1^2)\frac{M_{\tilde g}^2}{\alpha_3^2} +
4 m_W^2 -2m_Z^2, \;\;{\rm for\; {\bf 200}}.
\eea
From these sum rules we see that at the tree-level the average mass
squared difference between charginos and neutralinos is positive for
the representations {\bf 1}, {\bf 24} and {\bf 75}, whereas for the
representation {\bf 200} it is negative. In this respect the 
representation {\bf 200} resembles the anomaly
mediated supersymmetry breaking scenario, where it was found that the
average mass squared difference is negative \cite{HLP1}. In
Fig.~\ref{sumrule}, we have plotted the above sum rules for the
different $SU(5)$ representations that arise in (\ref{product}).  For
the numerical evaluation of the masses, we have used the program
SOFTSUSY \cite{SOFTSUSY}, including radiative corrections to the
neutralino and chargino masses.

\section{Higgs detection using $H^0,A^0 \rightarrow
  \tilde{\chi}_2^0\tilde{\chi}_2^0 \to 4 l$}

It is often assumed, when considering the detection of the Higgs
bosons in supersymmetric models, that supersymmetric partners are too
heavy so that Higgs bosons cannot decay into supersymmetric particles.
However, it may well be that for the heavy Higgs bosons $H^0$, $A^0$,
and $H^\pm$ the decays to supersymmetric particles are important or
even dominant \cite{cms1,cms2}. On the other hand, the decay branching 
ratios of neutralinos and charginos have been analyzed in \cite{BFM}.  
In the case of large $\tan\beta$, when the couplings to the heavy fermions 
are enhanced, the decays to the third generation particles have been 
discussed in \cite{ltanb}. For large values of $\tan\beta$, the decays 
to the third generation particles for the nonuniversal gaugino masses 
were discussed in~\cite{DMM}.  Here we are interested in Higgs decay to
$\tilde{\chi}_2^0$, which in turn decays to electrons and muons in the
case of non-universal gaugino masses.

\subsection{Decay of $\tilde{\chi}_2^0$ to leptons}

Of the supersymmetric particles, the light neutralinos
$\tilde{\chi}_{1,2}^0$, the light chargino $\tilde{\chi}_1^\pm$, and
the lightest sleptons are usually among the lightest particles in the
spectrum.  Higgs decays to sleptons are suppressed because of the
small coupling, which is proportional to the corresponding lepton
mass.  The decay to the lightest neutralino LSP is among the invisible
decays, which may be extremely difficult to detect at the LHC.  In the
minimal SUGRA model, the second lightest neutralino and the lighter of
the charginos have similar mass.  In \cite{cms1}, the decay of the
heavy neutral Higgs boson to a pair of the second lightest neutralinos
was studied.  It was found that in the case when the branching ratio
of $\tilde{\chi}_2^0$ to two leptons and the lightest neutralino is
large, then the possibilities of detection are promising.  Even though
the branching ratio to a chargino pair may be larger \cite{BBKT,moretti}, 
the decay to $\tilde{\chi}_2^0$'s is more promising because of the clear four
lepton signal.  We will, therefore, study the decay chain \bea H^0,
A^0\to\tilde{\chi}_2^0\tilde{\chi}_2^0,\;\; \tilde{\chi}_2^0\to
\tilde{\chi}_1^0 l^+l^-,\;\; l=e,\mu \eea for the four different
representations of $SU(5)$ in (\ref{product}). The decay
$\tilde{\chi}_2^0\to \tilde{\chi}_1^0 l^+l^-$ depends on the
parameters $M_2$, $M_1$, $\mu$, and $\tan\beta$, which control the
neutralino masses and the mixing parameters, and also on the slepton
masses $m_{\tilde l}$. As long as the direct decay of
$\tilde{\chi}^0_2$ into $\tilde{\chi}^0_1 + Z^0$ is suppressed and the
sleptons are heavier than the $\tilde{\chi}^0_2$, three body decays of
$\tilde{\chi}^0_2$ into charged leptons and $\tilde{\chi}^0_1$ will be
significant.  There can also be constructive or destructive
interference between the $Z^0$ and the slepton exchange amplitudes
which can have strong influence on the branching ratio. In some cases,
we also consider the possibility that the sleptons mediating the decay
$\tilde{\chi}_2^0\rightarrow \tilde{\chi}_1^0 l^+l^-$ can be on
mass-shell. Also, in order to have large branching ratios to the
sleptons, they should be lighter than the squarks. This is usually
true in SUSY models.
\begin{figure}
  \psfrag{sl}{\tiny${\tilde l}$}
  \includegraphics{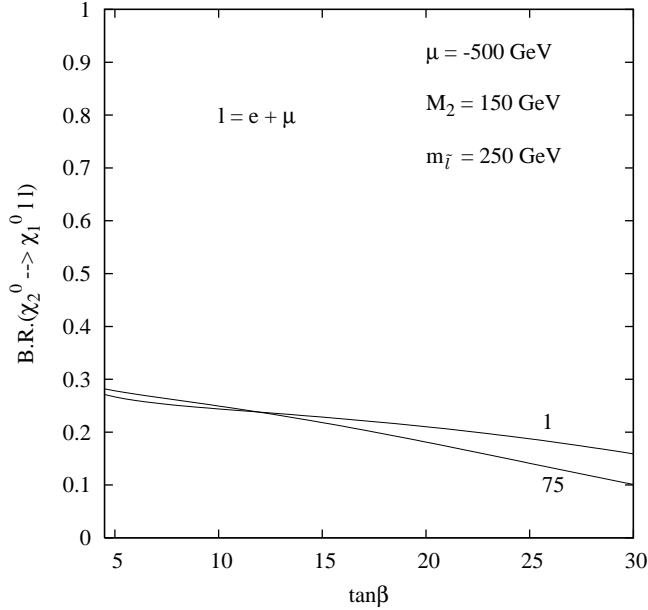} 
  \caption{\label{fig4} Branching ratio as a function of $\tan\beta$,
    in the case of $m_{\tilde l}$ $>$ $m_{\tilde{\chi}^0_2}$ and for
    the representations {\bf 1} and {\bf 75}.}
\end{figure}

In the singlet case, the three body decay $\tilde{\chi}_2^0\rightarrow
\tilde{\chi}_1^0 l^+l^-$ was discussed in Ref.~\cite{cms1} for a particular 
set of parameters, for which the branching ratio is large. Here, all the 
sleptons (including the staus) are assumed to have soft SUSY breaking 
masses of $250$ GeV and the value of $\mu$ = -500 GeV. In this analysis 
$M_2$ is a free parameter, and its value at the electroweak scale is taken 
to be 150 GeV. The squark masses are all taken equal to $1$ TeV. We have 
also taken a large value of the trilinear scalar coupling $A_t = 1$ TeV in 
order to have experimentally acceptable mass for the lightest Higgs boson. 
All the soft scalar masses, the value of $A_t$ and the value of $\mu$ are taken
at the electroweak scale. The pseudoscalar Higgs mass $m_A$ is a free 
parameter and its value is taken to be $340$ GeV.  The value of $M_1$ is 
determined from the ratio of the gaugino mass parameters in the singlet 
representation of $SU(5)$ in (\ref{product}). Due to the mentioned choice 
of $M_2$, $\tilde{\chi}^0_2$ is predominantly a wino, and $\tilde{\chi}^0_1$ 
is a bino-dominated state.  The decay of $\tilde{\chi}^0_2$ into
$\tilde{\chi}^0_1$ and a $Z^0$ is kinematically disallowed. The
branching ratio of the three-body decay is shown in Fig.~\ref{fig4} as
a function of $\tan\beta$ for the singlet case as well as for the
representation {\bf 75}. We have calculated the branching ratio using
the program SDECAY \cite{Muhlleitner:2003vg}.  In this figure the
initial value of $\tan\beta$ is 4.5, since for a lower value of
$\tan\beta$ the light Higgs mass $m_h$ is less than 114.4 GeV, which
is the LEP lower limit \cite{lephiggs}.  We see from the figure that
for higher values of $\tan\beta$ this branching ratio decreases since
the branching ratio $\tilde{\chi}^0_2 \to \tilde{\chi}^0_1 \tau^+
\tau^-$ increases with $\tan\beta$ due to a larger Yukawa coupling.

For the representation {\bf 75}, $\tilde{\chi}^0_2$ is wino-dominated
and $\tilde{\chi}^0_1$ is bino-dominated as in the singlet
case. However, the mass difference between the $\tilde{\chi}^0_2$ and
the $\tilde{\chi}^0_1$ is much smaller compared to the singlet
case. As we see from Fig.~\ref{fig4}, in the low $\tan\beta$ region
the branching ratio for these two different representations are very
close though the branching ratio for the {\bf 75} representation is
slightly larger. This is due to the fact that BR($\tilde{\chi}^0_2 \to
\tilde{\chi}^0_1 q {\bar q}$) is slightly larger in the singlet case
as compared to the case of {\bf 75} dimensional representation.  The
leptonic branching ratio is then almost equally distributed among the
available channels.  However, for large $\tan\beta$ the branching
ratio in the $\tilde{\chi}^0_1 \tau^+\tau^-$ channel is larger for the
{\bf 75} case than for the singlet case.  For large $\tan\beta$ this
makes the branching ratio in the $\tilde{\chi}^0_1 l^+l^-$ channel
smaller for the case of {\bf 75} dimensional representation.  We also
note that in the case of {\bf 75} dimensional representation the
partial decay width of $\tilde{\chi}^0_2 \rightarrow \tilde{\chi}^0_1
\nu \bar \nu$ is larger than the partial decay width of
$\tilde{\chi}^0_2 \rightarrow \tilde{\chi}^0_1 l^+ l^-$ in the large
$\tan\beta$ region. On the other hand, in the singlet case the partial
decay width of $\tilde{\chi}^0_2 \rightarrow \tilde{\chi}^0_1 \nu \bar
\nu$ is always smaller than that of $\tilde{\chi}^0_2 \rightarrow
\tilde{\chi}^0_1 l^+ l^-$.
\begin{figure}
  \psfrag{sl}{\tiny${\tilde l}$}
  \includegraphics{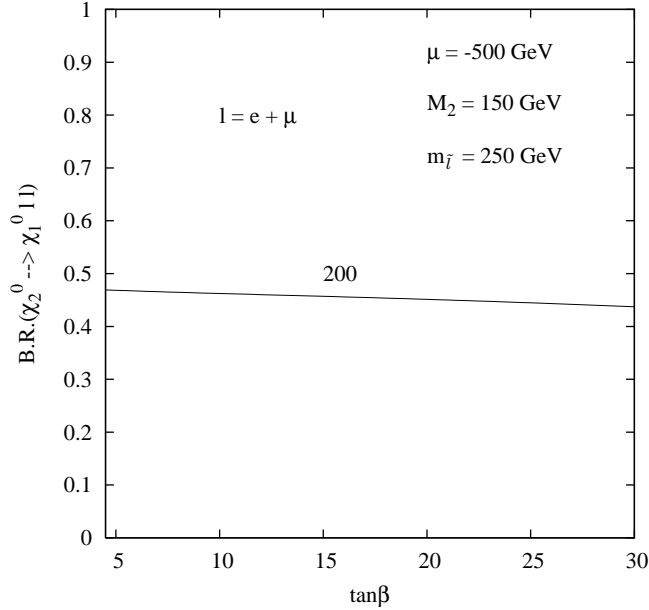}
  \caption{\label{fig5} Branching ratio as a function of $\tan\beta$,
    in the case of $m_{\tilde l}$ $<$ $m_{\tilde{\chi}^0_2}$, for the
    representation {\bf 200}.  All other parameters are same as in
    Fig.~\ref{fig4}.}
\end{figure}

For the set of parameters that we have discussed and in the case of the
representation {\bf 200}, the spectrum is such that all the left and
right handed sleptons are lighter than $\tilde{\chi}^0_2$ and are
produced on mass-shell. Although, $\tilde l_R$ proceeds with 100$\%$
branching ratio to $\tilde{\chi}^0_1 + l$, in the case of $\tilde l_L$
one should multiply by the appropriate branching fraction. This is
shown in Fig.~\ref{fig5}.  One can see that the dependence on
$\tan\beta$ is not significant in this case.  The results for the
representation {\bf 24} are not shown here due to the fact that it
results in the lightest neutralino mass below the current experimental
lower limit.

\subsubsection{The case of {\bf 200}}
In this subsection we will consider the representation {\bf 200} of
$SU(5)$, as it arises in (\ref{product}), in some detail. The ratio of
the $U(1)$ and the $SU(2)$ gaugino masses is approximately given by
(at the one-loop level\footnote{Using the two-loop renormalization
group equations (RGEs) the ratio is approximately 2.6:1 for a wide
range of parameter choices. Similarly, for other representations the
change in this ratio with the use of two-loop RGEs does not change our
conclusions.}) \bea |M_1| : |M_2| = 2.5 : 1. \eea This resembles very
much the scenario of anomaly mediated supersymmetry breaking for
values of $\mu$ larger than $M_2$. Let us now highlight two important
characteristics of this representation.

$\bullet$ ${\tilde{\chi}}_1^\pm$ and ${\tilde{\chi}}_1^0$ are almost
exclusively winos, and they are nearly degenerate in mass.

$\bullet$ ${ \tilde{\chi}}_2^0$ is predominantly a bino for $|\mu| > M_1$.

\begin{figure}
  \psfrag{sl-L}{\tiny${\tilde l}_L$}
  \psfrag{sl-R}{\tiny${\tilde l}_R$}
  \includegraphics{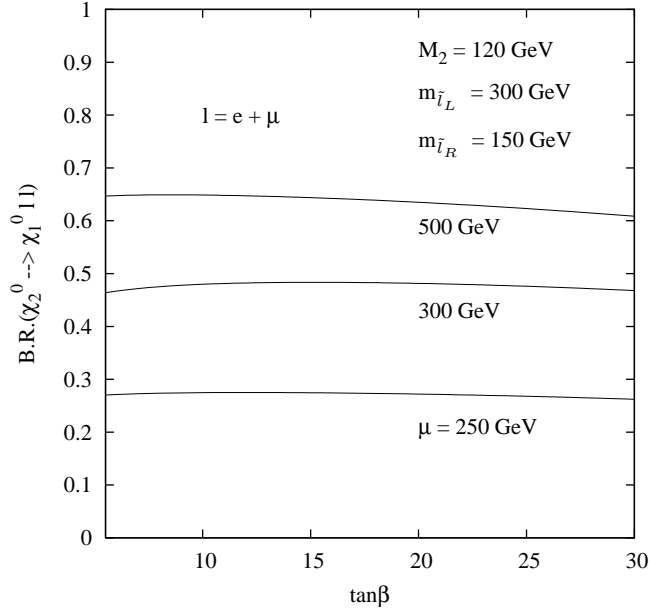}
  \caption{\label{fig6} Branching ratio as a function of $\tan\beta$,
    in the case of $m_{{\tilde l}_R}$ $<$ $m_{\tilde{\chi}^0_2}$ and
    for the representation {\bf 200} and for three different values of
    $\mu$.}
\end{figure}

\noindent Consider the decay ${\tilde{\chi}}_2^0\rightarrow
{\tilde{\chi}}_1^0 l^+ l^-$ for the ${\bf 200}$ dimensional
representation. We again choose the scalar masses in such a way that
$m_{{\tilde l}_R},m_{{\tilde \tau}_1}$ $<$ $m_{\tilde{\chi}^0_2}$ $<$
$m_{{\tilde l}_L}$, $m_{\tilde \nu}$. The reason for such a choice is
that the two-body decay ${\tilde{\chi}}_2^0 \rightarrow {\tilde l}_R
l$ is allowed. Although the decay of ${\tilde l}_R$ into $l +
{\tilde{\chi}}_1^0$ is highly suppressed due to the very small bino
component in ${ \tilde{\chi}}_1^0$, ${\tilde l}_R$ will decay
eventually in this mode with a one hundred percent branching ratio. Of
course, one should be careful to consider the possibility of a
displaced vertex in the decay of ${\tilde l}_R$. The
BR(${\tilde{\chi}}_2^0 \rightarrow {\tilde{\chi}}_1^0 l^+ l^-$)
calculated in this manner depends very strongly on $\mu$ (increases
with increasing $\mu$) since as $\mu$ increases ${\tilde \chi}_2^0$
becomes more and more bino like, and thus the partial decay widths of
${\tilde{\chi}}_2^0 \rightarrow {\tilde{\chi}}_1^\pm W^\mp$ and ${
\tilde{\chi}}_2^0 \rightarrow { \tilde{\chi}}_1^0 h$ are suppressed
and the partial decay width of ${\tilde{\chi}}_2^0 \rightarrow {\tilde
l}_R l$ is enhanced. This makes the branching ratio into ${\tilde l}_R
l$ mode larger for large values of $\mu$. This is shown in
Fig.\ref{fig6}. The branching ratio in the channel ${\tilde{\chi}}_2^0
\rightarrow {\tilde{\chi}}_1^0 Z$ is always very small for all values
of $\tan\beta$.

We note that in the case of the ${\bf 200}$ dimensional representation
we use the constraint $m_{{\tilde{\chi}}_1^\pm}$ $>$ 88 GeV applicable
for nearly mass degenerate lighter chargino and the lightest
neutralino \cite{alephdegn}.

\begin{figure}
  \psfrag{sl-L}{\tiny${\tilde l}_L$}
  \psfrag{sl-R}{\tiny${\tilde l}_R$}
  \includegraphics{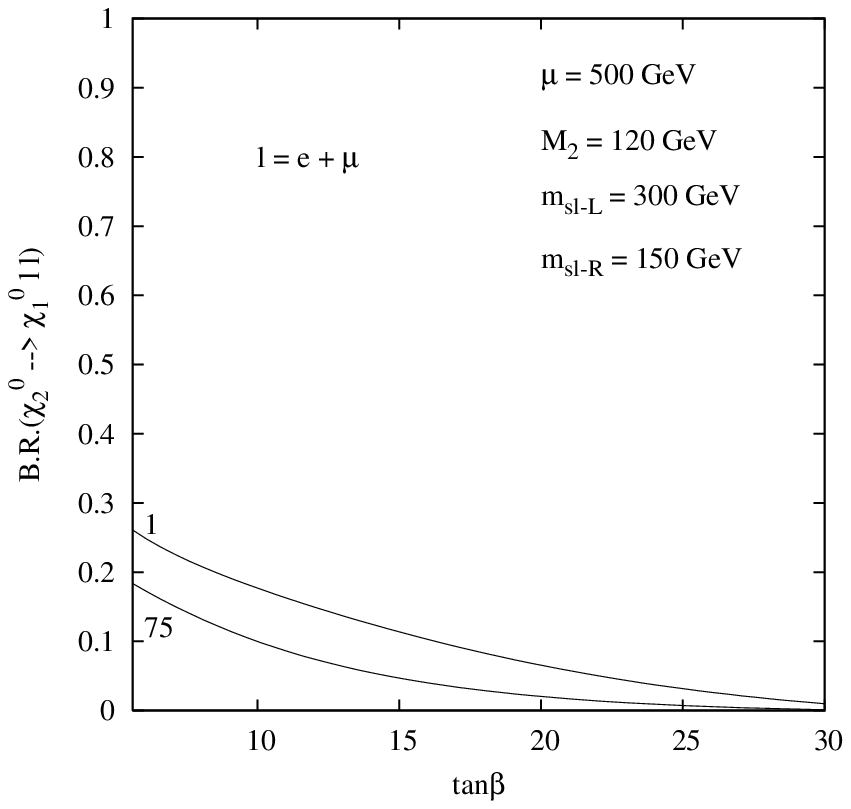}
  \caption{\label{fig7} Branching ratio as a function of $\tan\beta$,
    in the case of $m_{{\tilde l}}$ $>$ $m_{\tilde{\chi}^0_2}$ and for
    the representations {\bf 1} and {\bf 75} and for $\mu$ = 500 GeV.
    All other parameters are the same as in Fig.~\ref{fig6}.}
\end{figure}

Let us now discuss the BR(${ \tilde{\chi}}_2^0 \rightarrow
{\tilde{\chi}}_1^0 l^+ l^-$) for this set of parameters for the
representations ${\bf 1}$ and ${\bf 75}$. We do not compare the case
for the representation ${\bf 24}$ here since for the parameter choice
of this figure the {\bf 24} dimensional representation always produces
a lightest neutralino with mass below the current experimental
limit \cite{PRD}. For this set of parameters the representations ${\bf 1}$ 
and ${\bf 75}$ give similar kind of spectrum so that no two-body decays of
${\tilde{\chi}}_2^0$ are allowed. In Fig.~\ref{fig7} we show the
branching ratio for these two representations as a function of
$\tan\beta$ and for a value of $\mu$ = 500 GeV.

\subsubsection{The case of {\bf 24}}

\begin{figure}
  \psfrag{sl-L}{\tiny${\tilde l}_L$}
  \psfrag{sl-R}{\tiny${\tilde l}_R$}
  \includegraphics{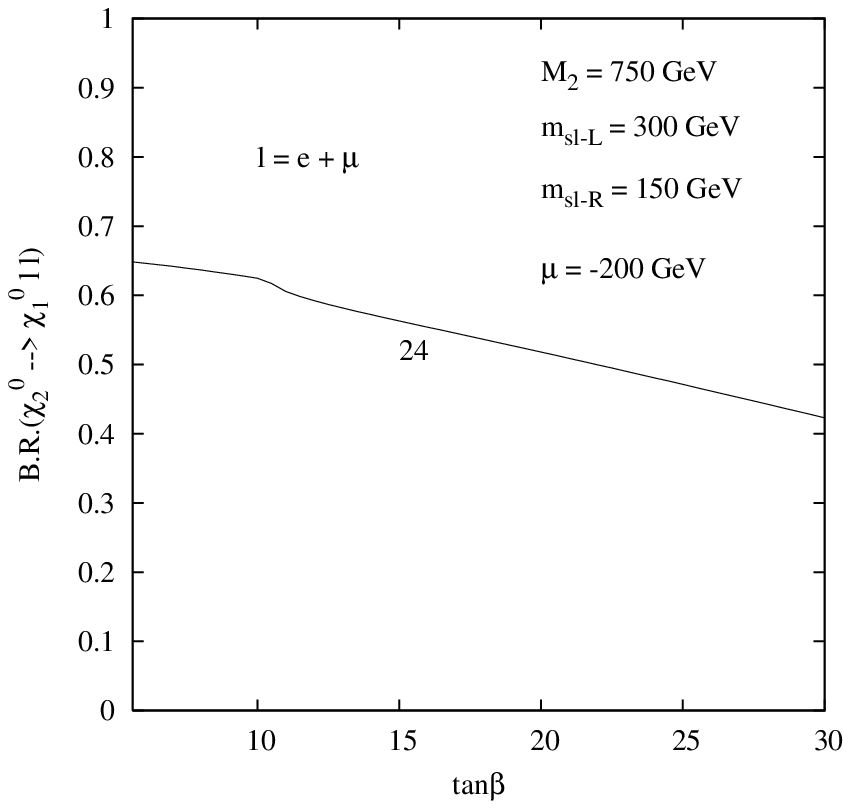}
  \caption{\label{fig8} Branching ratio as a function of $\tan\beta$,
    in the case of $m_{{\tilde l}_R}$ $<$ $m_{\chi^0_2}$ and for the
    representation {\bf 24}. The choice of other parameters is
    described in the text.}
\end{figure}

In this subsection we will consider the case of ${\bf 24}$ dimensional
representation where $|M_1| \approx 0.166 |M_2|$. We look for a set of
parameters such that the mass of the lightest neutralino is not below
the current experimental lower limit as was the case in the previous
sub-sections. For the present study we again consider the mass
spectrum $m_{{\tilde \chi}_1^0} < m_{{\tilde l}_R} < m_{{\tilde
\chi}_2^0}$. We have taken $M_2$ = 750 GeV and $\mu$ = -200 GeV. For
this choice of the parameters the lightest neutralino ${\tilde
\chi}_1^0$ is mostly a bino with some higgsino admixture whereas the
second lightest neutralino is higgsino dominated with very small wino
and bino components. The soft masses for the left-sleptons are assumed
to be 300 GeV, whereas those of the right-sleptons are taken to be 150
GeV.  Other parameters such as squark masses and the trilinear scalar
coupling $A_t$ are same as before. Once again we have taken the values
of these parameters at the electroweak scale. With this set of parameters 
the following two-body decay channels are dominant : ${\tilde \chi}_2^0
\rightarrow {\tilde l}_R l$ and ${\tilde \chi}_2^0 \rightarrow {\tilde
\tau}_1 \tau$. We have plotted the branching ratio of ${\tilde
\chi}_2^0 \rightarrow {\tilde \chi}_1^0 l^+ l^-$ as a function of
$\tan\beta$ in Fig.~\ref{fig8}. We see from this Fig.~ that for some
values in the low $\tan\beta$ region this branching ratio can be as
large as $65\%$.  For large values of $\tan\beta$, the BR(${\tilde
\chi}_2^0 \rightarrow {\tilde \chi}_1^0 \tau^+ \tau^-$)
dominates. Also, for this choice of parameters in the {\bf 1}, {\bf
75} and {\bf 200} representations in (\ref{product}) we always have a
stau LSP which we do not consider in our R-parity conserving scenario.

\subsection{Decay of  heavy Higgs bosons into a pair of neutralinos 
  : $H^0,A^0 \rightarrow \tilde{\chi}_2^0\tilde{\chi}_2^0$}

In this subsection we will study the branching ratios of the heavy
Higgs bosons $H^0$ and $A^0$ into a pair of second lightest
neutralinos. We have used the package HDECAY \cite{hdecay} to
calculate the branching ratios.  The decay widths and the branching
ratios depend on the ratio of $M_1$ and $M_2$ along with other MSSM
parameters. We have calculated the branching ratio of $H^0,A^0
\rightarrow \tilde{\chi}_2^0\tilde{\chi}_2^0$ for different $SU(5)$ 
representations that arise in the product (\ref{product}).
The coupling of the heavy Higgs boson $H^0$ with a pair of neutralinos
is given by \cite{gunion-haber,higgs-hunters}: \bea
H^0{{\tilde{\chi}}^0_i { \tilde{\chi}}^0_j} :~~ -{ig}(A_L P_L + A_R
P_R), \eea
\noindent 
where $P_L = {1 \over 2}(1-\gamma_5)$ and $P_R = {1 \over 2}(1+\gamma_5)$
are the usual projection operators. The coefficients of $P_L$ and $P_R$ are
given by  
\bea
A_L = Q^{{\prime\prime}*}_{ji} \cos\alpha -
S^{{\prime\prime}*}_{ji} \sin\alpha ,\;\;
A_R = Q^{{\prime\prime}}_{ij} \cos\alpha - S^{{\prime\prime}}_{ij}
\sin\alpha, 
\eea 
where
\bea
Q^{\prime\prime}_{ij} = {1 \over 2}[Z_{i3}(Z_{j2} -Z_{j1}\tan\theta_W) +
Z_{j3}(Z_{i2} -Z_{i1}\tan\theta_W)]\epsilon_i,  \nonumber
                        \\[1.2ex]
S^{\prime\prime}_{ij} = {1 \over 2}[Z_{i4}(Z_{j2} -Z_{j1}\tan\theta_W) +
Z_{j4}(Z_{i2} -Z_{i1}\tan\theta_W)]\epsilon_i.  
\eea 
\noindent Here $Z$ is the neutralino mixing matrix in the basis
$(-i{\tilde B}, -i{\tilde W}, {\tilde H}_1, {\tilde H}_2)$, and
$\epsilon_i$ is the sign of the i$^{th}$ neutralino mass eigenvalue.
Furthermore, $\sin\alpha$ and $\cos\alpha$ are the usual Higgs mixing
angles.

Similarly, the coupling of the pseudoscalar Higgs boson $A^0$ to a
pair of neutralinos is given by \cite{gunion-haber,higgs-hunters} :
\bea A^0{{\tilde{\chi}}^0_i {\tilde{\chi}}^0_j} :~~ -{g}(B_L P_L -
B_R P_R), \eea
\noindent where the coefficients of $P_L$ and $P_R$ are
given by
\bea
B_L = Q^{{\prime\prime}*}_{ji} \sin\beta -
S^{{\prime\prime}*}_{ji} \cos\beta ,\;\;
B_R = Q^{{\prime\prime}}_{ij} \sin\beta - S^{{\prime\prime}}_{ij}
\cos\beta.
\eea

\begin{figure}
  \psfrag{sl-L}{\tiny${\tilde l}_L$}
  \psfrag{sl-R}{\tiny${\tilde l}_R$}
  \includegraphics{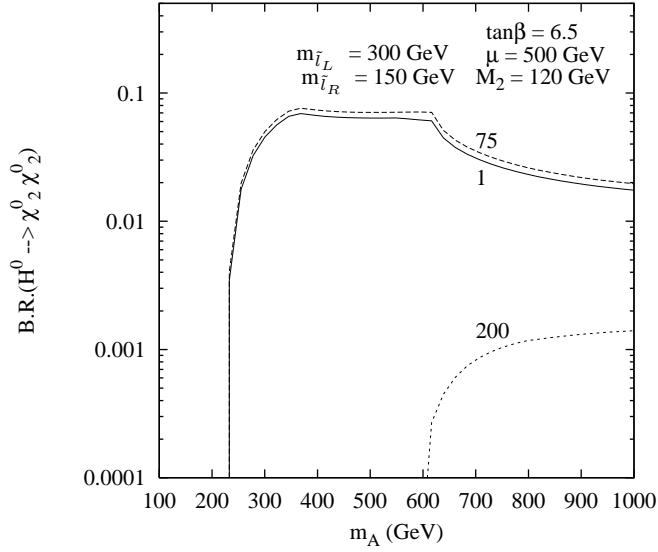}
  \caption{\label{fig9} The branching ratio of $H^0 \rightarrow
    \tilde{\chi}_2^0\tilde{\chi}_2^0$ as a function of $m_A$ for three
    different $SU(5)$ representations in (\ref{product}). Here
    $\tan\beta$ is taken to be 6.5, and other MSSM parameters are same
    as in Fig.~\ref{fig7}.}
\end{figure}

\begin{figure}
  \psfrag{sl-L}{\tiny${\tilde l}_L$}
  \psfrag{sl-R}{\tiny${\tilde l}_R$}
  \includegraphics{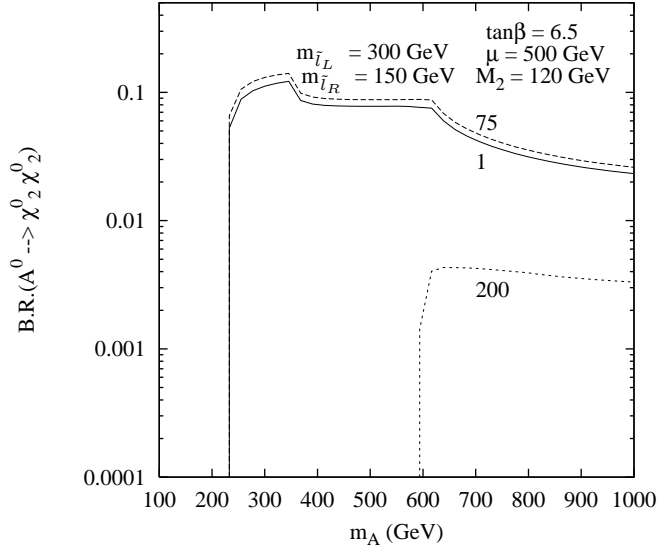}
  \caption{\label{fig10} The branching ratio of $A^0 \rightarrow
    \tilde{\chi}_2^0\tilde{\chi}_2^0$ as a function of $m_A$ for three
    different $SU(5)$ representations.  Here, $\tan\beta$ is taken to
    be 6.5, and other MSSM parameters are same as in Fig.~\ref{fig9}.}
\end{figure}

As an example, in Fig.~\ref{fig9}, we have shown the dependence of
branching ratio BR($H^0 \rightarrow \tilde{\chi}_2^0\tilde{\chi}_2^0$)
on $m_A$ for a particular choice of MSSM parameters. This point in the
parameter space is the same as in Fig.~\ref{fig7} with the choice of
$\tan\beta$ = 6.5. This way we can directly compare the branching
ratios in the representations ${\bf 1}$, ${\bf 75}$ and ${\bf
200}$. In Fig.~ \ref{fig10}, we have plotted the branching ratio
BR($A^0 \rightarrow \tilde{\chi}_2^0\tilde{\chi}_2^0$) as a function
of $m_A$ for the same choice of parameters, and for the same $SU(5)$
representations. We can see that for this choice of the parameter set
and for $m_A <$ 350 GeV, the branching ratio of the decay of $A^0$ is
larger than that of the decay of the heavy Higgs scalar $H^0$ for the
representations ${\bf 1}$ and ${\bf 75}$. This is due to the fact that
for $H^0$ the total decay width is larger due to the increase in the
number of available channels to the SM particles, which leads to a
smaller branching ratio to sparticles. In the case of ${\bf 200}$
dimensional representation the threshold opens up for heavier $m_A$,
and once again the branching ratio of $A^0$ is larger than that of the
$H^0$.  As we have discussed earlier, the representation ${\bf 24}$
produces a very light lightest neutralino for this choice of
parameters, and is not further discussed here.

\subsection{Signal cross section}

We now consider signal cross section and the total event rate in the
four lepton channel at the LHC with $\sqrt s$ = 14 TeV for two
different representation, the singlet and ${\bf 75}$.  We show the
contours of constant cross section in the ($m_A$, $\tan\beta$) plane
for a representative set of MSSM parameters at the electroweak scale. As 
in Fig.~\ref{fig9}, we have taken $\mu = 500$ GeV, $M_2 = 120$ GeV, all 
left slepton masses to be 300 GeV and all the right slepton masses to be 
150 GeV. All squark masses are taken to be 1 TeV. The top mass is $m_t$ 
= 178 GeV and the bottom mass $m_b$ is 4.25 GeV.
\begin{figure}
  \includegraphics{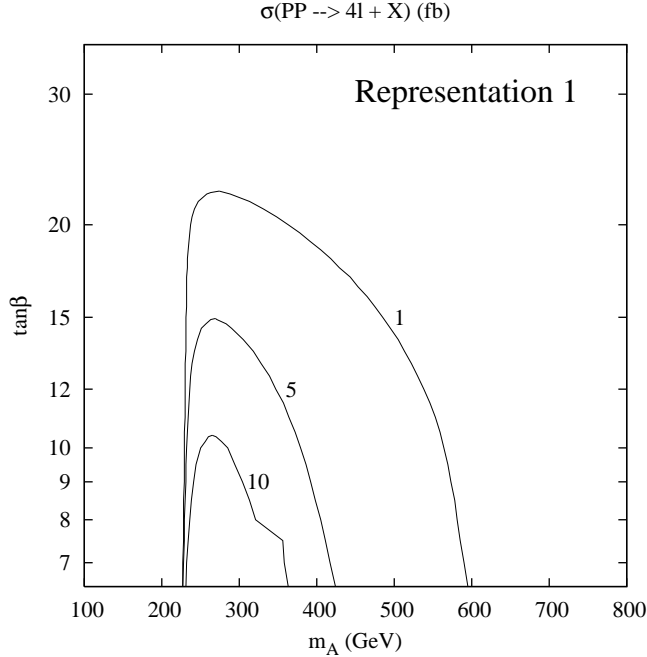}
  \caption{\label{fig11} Contours of $\sigma(pp \longrightarrow H^0,A^0
    \longrightarrow \tilde{\chi}_2^0\tilde{\chi}_2^0 \longrightarrow 4
    l + X)$ in fb, where $l = e^\pm$ or $\mu^\pm$ and $X$ represents
    invisible final state particles. This is the case for the singlet
    representation.  Other MSSM parameters are the same as in
    Fig.~\ref{fig9}. $\sqrt s$ = 14 TeV.}
\end{figure}
\noindent The production cross section $gg \rightarrow H^0/A^0$ has
been calculated in the next-to-leading order using the package HIGLU
\cite{gghspira}, which is based on the calculations in
Ref.\cite{SDGZ}. For the gluon distribution function we have used the
distribution given in \cite{GRV}. We note that for low values 
of $\tan\beta$ this
channel dominates the production cross section. We have also
considered the inclusive associated production $q {\bar q}, gg \rightarrow b
{\bar b} H^0/A^0$ at the leading order \cite{hbbar} (which is
essentially the leading order subprocess $b \bar b \rightarrow H^0/A^0$). 
The factorization and the renormalization scale are chosen to be $\mu_F$
= $\mu_R$ = $(m_{H/A} + 2 m_b)/2$.  The process $gg \rightarrow 
b {\bar b} H^0/A^0$  as well as the process  $gg \rightarrow H^0/A^0$
is enhanced for larger values of $\tan\beta$ due to the large coupling of Higgs 
bosons to $b {\bar b}$. However, the process $gg \rightarrow
b {\bar b} H^0/A^0$  dominates the production process for the 
values of $m_A$ that we have considered here ($m_A \gsim$ 200 GeV). In order 
to get a quantitative idea of these individual contributions to the 
Higgs boson production let us give an example here. If we take 
$\tan\beta$ = 20 and $m_A$ = 230 GeV then $gg \rightarrow H$ 
production cross section is 4.08 pb and $Hb \bar b$ production 
cross section is 31.4 pb at LHC energy.
\begin{figure}
  \includegraphics{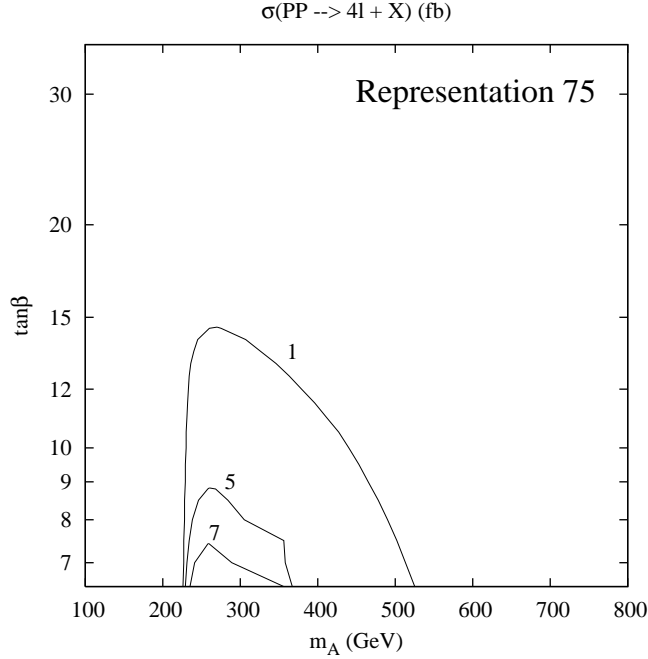}
  \caption{\label{fig12} Contours of $\sigma(pp \longrightarrow H^0,A^0
    \longrightarrow \tilde{\chi}_2^0\tilde{\chi}_2^0 \longrightarrow 4
    l + X)$ in fb, where $l = e^\pm$ or $\mu^\pm$ and $X$ represents
    invisible final state particles. This is the case for the
    representation ${\bf 75}$.  Other MSSM parameters are the same as
    in Fig.~\ref{fig9}. $\sqrt s$ = 14 TeV.}
\end{figure}
Next, we multiply these Higgs production cross sections by the
appropriate branching ratios B.R.($H^0/A^0 \to \tilde \chi^0_2 
\tilde \chi^0_2$) and B.R.($\tilde \chi^0_2 \to \tilde \chi^0_1 l^+l^-$)
discussed in the previous subsections in order to get the four lepton
signal at the LHC. In Fig.~\ref{fig11} we have shown the contours of constant 
cross section of the 4l signal for the singlet representation arising in 
(\ref{product}). We see that for $\tan\beta$ up to $\approx$ 10, and 
$m_A \sim$ 250-350 GeV the total $4l$ cross section can reach up to 10 fb, 
which corresponds to 1000 signal events (without any cuts) for integrated
luminosity of 100 fb$^{-1}$. In Fig.~\ref{fig12}, the contours of
constant cross section are shown for the representation ${\bf 75}$ for
the same choice of parameters. It is evident from this figure that a
smaller region in the ($m_A$, $\tan\beta$) plane can be probed in this
case with the same number of events. However, it shows different
possibilities for these two representations. It is also evident from these 
two figures that the four lepton signal is very small for large values
of $\tan\beta$ due to the suppression of the branching ratio
B.R.($H^0/A^0 \to \tilde \chi^0_2 \tilde \chi^0_2$). In the case of
the representation ${\bf 200}$, this four lepton signal is available
only for higher values of $m_A$ ($>$ 600 GeV) as can be seen from Fig.
\ref{fig9} and Fig. \ref{fig10} for identical choices of other
parameters. However, the total cross section is less than 1 fb for a 
large region in the ($m_A$--$\tan\beta$) plane and we do not show any
separate plot for that. The total cross section in this four lepton 
channel can be similarly studied for the representation ${\bf 24}$ for
some different set of parameters which we do not pursue here.

We now briefly discuss the possible backgrounds to this four lepton
signal. There can be two types of backgrounds, namely the Standard
Model processes leading to this type of signal, and SUSY
processes. The main SM background comes from $Z^0Z^0$ and $t {\bar t}$
production.  As discussed in Ref. \cite{cms1}, the $t {\bar t}$
background can be eliminated to a large extent by requiring four
isolated leptons with $p^T_l >$ 10 GeV.  Demanding a missing
transverse energy of 20 GeV and an explicit $Z^0$ veto can reduce the
background from $Z^0Z^0$.  The background from SUSY processes can come
from squark/gluino production or sneutrino pair production. The events
coming from squark/gluino production can be eliminated by requiring
soft jets with $E_T <$ 100 GeV and $E^{miss}_T <$ 130 GeV. The
background coming from sneutrino pair production is more difficult to
handle. However, it could possibly be distinguished due to the fact
that it has larger $E^{miss}_T$ and larger $p^T_l$ compared to the
signal. After using all these cuts the percentage of four lepton
signal surviving is approximately about 60$\%$. A detailed simulation
of the signal and background events is beyond the scope of the present
work.  We hope to come back to this issue in a future work.

\section{Higgs production in the cascade $\tilde q,\tilde
  g\to\tilde{\chi}_2^0 \to h\tilde{\chi}_1^0$}

If squarks and gluinos are light enough to be produced ($pp \to \tilde q
\tilde q^\prime, \tilde g \tilde g, \tilde g \tilde q$), then their
production cross section will be large at a hadron collider.  Thus,
the decay chain \bea \tilde q,\tilde g\to \tilde{\chi}_2^0
+X \to \tilde{\chi}_1^0 h (H^0,A^0)+X \to \tilde{\chi}_1^0 b\bar
b + X \label{chain} \eea will be an important source to look for Higgs 
bosons at LHC in the final state $b \bar b b \bar b$ + X.
This chain has been considered at LHC with universal gaugino masses in
\cite{DDGM,cms2}.  It was found that for suitable values of the
parameters, the signal for all of the neutral Higgs bosons was clearly
above the background.  Discovery potential for $A^0$ and $H^0$
extended to 200 GeV, independent of the values of $\tan\beta$.
 
Here we will consider the decay chain in (\ref{chain}) for
nonuniversal gaugino masses, and study the changes that occur from the
case of universal gaugino masses. We will only consider the cases
without top squarks, which is enough to illustrate the differences
that arise when the gaugino masses are nonuniversal. 
 
In proton collisions, a squark pair, squark-antisquark, squark-gluino,
or a gluino pair can be produced. We have used PROSPINO
\cite{Beenakker:1996ed} to calculate the production cross sections of
these modes (for $\sqrt{s}=14$ TeV).  Here we study the part of the 
parameter space for which $m_{\tilde g}>m_{\tilde q}$. Then every gluino 
decays to a quark and the corresponding squark. If the mass difference 
of the (top) squark and gluino is large enough then gluino may also decay 
to a top-stop pair. Because we are considering only the five lightest 
quark flavors, we have to remove top-stop contribution from the results.

In order to compare with the calculation in the universal (singlet)
case in \cite{DDGM}, we use the average branching ratios for
particles, as used in \cite{DDGM}, and which are defined in \cite{datta}.
Thus we sum over all the decay widths of squarks decaying into a quark
and a neutralino divided by the total decay width of squarks decaying
into any neutralino and quark. The five squarks (excluding the stop) are 
considered to be equal in the sense that the bottom Yukawa coupling effect is
neglected. The decay branching ratios are calculated using SDECAY
\cite{Muhlleitner:2003vg} for squarks and neutralinos, and HDECAY
\cite{hdecay} for the final decay of Higgs bosons to bottom quarks.
 
\begin{figure}
  \psfrag{mg}{$m_{\tilde g}$ \scriptsize[GeV]}
  \psfrag{cs}{\footnotesize$\sigma (p p\to 4b + X)$ [pb]}
  \centering \includegraphics[width=7.5cm]{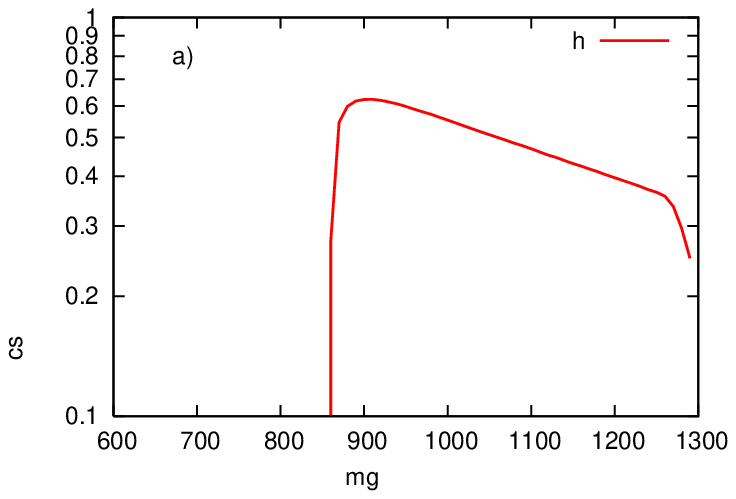}
  \centering \includegraphics[width=7.5cm]{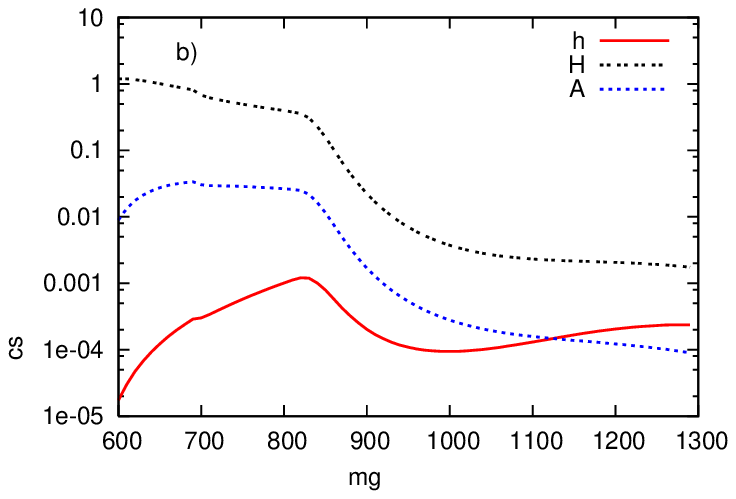}
  \caption{\label{figCSHA1}Cross section $p p\to b \bar b b \bar b$ +X
   at LHC through the decay chain $\tilde q,\tilde g\to
    \tilde{\chi}_2^0\to \tilde{\chi}_1^0 h (H^0,A^0)\to \tilde{\chi}_1^0
    b\bar b + X$ for (a) the
    representation~{\bf 1} via $h$, and (b) the representation~{\bf
    24} via $h, H^0$ and $A^0$.}  \hfill
\end{figure}

In Fig.~\ref{figCSHA1} (a) we have plotted the cross section of the
decay chain (\ref{chain}) as a function of the gluino mass for the
singlet representation. We have used parameters $\tan \beta = 10,
m_A=200$ GeV, $\mu=+500$ GeV, $m_{\tilde q}=600$ GeV and $m_{\tilde
l}=350$ GeV as low scale input values.
 
In the case of singlet representation, only the decay through the
light Higgs boson $h^0$ is kinematically possible. The mass difference
of the two lightest neutralinos is too small to produce heavier Higgs
bosons $H^0$ and $A^0$.  We can see a sharp rise in the cross section
where the light Higgs channel opens up. This is due to an increase in 
the $\tilde{\chi}_2^0$ and $\tilde{\chi}_1^0$ mass difference as a
function of the gluino mass. The production cross section of squarks
decreases as gluino mass increases. This is of course independent of
the representations arising in the product~(\ref{product}).
 
In Fig.~\ref{figCSHA1} (b) we have plotted the corresponding cross
section for the {\bf 24} dimensional representation. Because of the
changed relations between the gaugino mass parameters, the composition
and masses of the neutralinos are different from the universal
case. Now all the Higgs channels are available. We see that the $CP$-even
neutral Higgs $H^0$-channel gives the largest cross section.  As
gluino mass increases, the decay branching ratio of
$\tilde{\chi}_2^0\to h,H^0,A^0 + \tilde{\chi}_1^0$ decreases.
 
In the {\bf 75} dimensional representation the mass difference of the
two lightest neutralinos are generally too small in order for
$\tilde{\chi}_2^0$ to decay into Higgs bosons, thereby making the
decay chain, Eq.~(\ref{chain}), irrelevant.

For the {\bf 200} dimensional representation, the mass difference of
the two lightest neutralinos depends on the squark masses. Requiring
the lightest neutralino to be the LSP, and for experimentally viable
Higgs boson mass, the mass difference of the two lightest neutralinos
is relatively small, and the total cross section resulting from the 
decay chain (\ref{chain}) remains below the detection level.
 
\section{Conclusions}

We have studied the consequences of gaugino mass nonuniversality as it
arises in a supersymmetric grand unified theory for neutralino masses
and mass relations, as well as for particular Higgs production and
decay processes.

We found that the upper bounds of neutralino masses and the mass sum 
rules depend significantly on the representation.  
Similarly the studied decay possibilities of
Higgs bosons depend on the representations. The decay of
the second lightest neutralino to two leptons and the lightest
neutralino very much depends on the mass difference between the lightest 
and second lightest neutralino which in turn depends on the
representations. This is also true for the production of 
Higgs bosons in the decay of the second lightest neutralino.

{}From our considerations it seems clear that depending on the region
of the parameter space, the Higgs decay $h (H^0,A^0)\rightarrow
\tilde{\chi}_2^0\tilde{\chi}_2^0$ may be observable for the gauginos
emerging in any of the representations {\bf 1}, {\bf 24}, {\bf 75}, or
{\bf 200} of $SU(5)$.  However, the region in which $\tilde{\chi}_2^0
\rightarrow 2l+X$ is large, and it is possible for Higgs bosons to
decay to the second lightest neutralinos, is rather limited in any of
these models.  Thus, the relevant regions of the parameter space do
not necessarily overlap.  In Section 3 we compared singlet and {\bf
75} using the same set of parameters. For {\bf 200}, the total cross
section is less than 1 fb for the same choice of parameters when
kinematically available. For the representation ${\bf 24}$, we did not
find a region where a comparison could have been made. 

Interestingly, for the production of the Higgs bosons via the decay
chain including $\tilde{\chi}_2^0\rightarrow h (H^0,A^0)
\tilde{\chi}_1^0$, in addition to the singlet, we found relevant
region of the parameter space only for the representation {\bf 24}.
Furthermore, in this region the signal cross section for both neutral
heavy Higgs bosons is reasonably large for not very heavy gluinos.  It
should be noted that in these two cases the signatures are clearly
different. In the representation {\bf 24} the cross section is largest
at the lighter values of the gluino mass for the heavy Higgses. For
the lightest Higgs the production channel is open for all of the
discussed gluino masses in the representation {\bf 24} as opposed to
the case of {\bf 1} representation.  Also the fact that all the
neutral Higgs channels are open in the {\bf 24} case distinguishes it
from the singlet case, where only the light Higgs channel is
available.

Finally, we note that it is possible to find similar signatures from
the scenarios with non-universal Higgs masses \cite{Baer:2005bu}.

\vspace*{0.7cm}
\noindent
{\bf Acknowledgments}

One of the authors~(PNP) would like to thank the Helsinki Institute of
Physics, where this work was initiated, for its hospitality.  This
work was supported by the Academy of Finland (Project numbers 104368
and 54023), and by the Council of Scientific and Industrial Research,
India.\\

\end{document}